\documentclass[10pt,conference]{IEEEtran}
\usepackage{graphicx}
\usepackage{amssymb,amsmath,mathrsfs}
\usepackage{verbatim}

\begin{document}

\title{Capacity and Normalized Optimal Detection Error in Gaussian Channels}
\author{
\IEEEauthorblockN{Edwin Hammerich}
\IEEEauthorblockA{Ministry of Defence, 95030 Hof, Germany\\
E-mail: edwin.hammerich@ieee.org}
}
\maketitle

\begin{abstract}
For vector Gaussian channels, a precise differential connection between channel capacity and a quantity termed 
normalized optimal detection error (NODE) is presented. Then, this C--NODE relationship is extended to continuous-time 
Gaussian channels drawing on a waterfilling characterization recently found for the capacity of continuous-time 
linear time-varying channels. In the latter case, the C--NODE relationship becomes asymptotic in nature. In 
either case, the C--NODE relationship is compared with the I--MMSE relationship due to Guo et al.\ connecting 
mutual information in Gaussian channels with the minimum mean-square error (MMSE) of estimation theory.
\end{abstract}

\newtheorem{definition}{Definition}
\newtheorem{theorem}{Theorem}
\newtheorem{lemma}{Lemma}
\newtheorem{example}{Example}
\newtheorem{remark}{Remark}

\newcommand{\g}{\boldsymbol{g}}
\newcommand{\f}{\boldsymbol{f}}
\newcommand{\n}{\boldsymbol{n}}
\newcommand{\x}{\boldsymbol{x}}
\newcommand{\y}{\boldsymbol{y}}
\renewcommand{\u}{\boldsymbol{u}}
\renewcommand{\v}{\boldsymbol{v}}
\newcommand{\X}{\boldsymbol{X}}
\newcommand{\Y}{\boldsymbol{Y}}
\newcommand{\Z}{\boldsymbol{Z}}
\newcommand{\N}{\boldsymbol{N}}
\newcommand{\F}{\boldsymbol{F}}
\newcommand{\G}{\boldsymbol{G}}
\newcommand{\MatH}{\boldsymbol{H}}
\newcommand{\DELTA}{\boldsymbol{\Delta}}
\newcommand{\SIGMA}{\boldsymbol{\Sigma}}
\newcommand{\diag}{\mathrm{diag}}
\newcommand{\T}{\mathsf{T}}
\newcommand{\tr}{\mathrm{tr}}
\newcommand{\E}{\mathsf{E}}
\renewcommand{\S}{\mathscr{S}}
\renewcommand{\d}{\mathrm{d}}
\newcommand{\snr}{\mathrm{snr}}
\newcommand{\node}{\mathrm{node}}
\newcommand{\mmse}{\mathrm{mmse}}
\renewcommand{\P}{\boldsymbol{P}}
\newcommand{\R}{\mathbb{R}}

\section{Introduction}
The central result of Guo et al.\ in \cite{GSV} is an identity connecting mutual information in Gaussian channels 
with the MMSE of estimation theory. This I--MMSE relationship reads in the case of a vector Gaussian channel
\begin{equation}
  \frac{\d}{\d\,\snr} I(\boldsymbol{X};\sqrt{\snr}\boldsymbol{H}\boldsymbol{X}+\boldsymbol{N})=\frac{1}{2}
                                           \mmse(\snr),                                   \label{I-MMSE}
\end{equation}
where $\snr\ge0$, $\boldsymbol{N}$ is a noise vector with independent standard Gaussian components, independent of 
the random vector $\boldsymbol{X}$, $\E\|\boldsymbol{X}\|^2<\infty$, and $\boldsymbol{H}$ is a deterministic matrix 
of appropriate dimension; $\mmse(\snr)$ is the MMSE in estimating $\MatH\X$ given
\begin{equation}
  \Y=\sqrt{\snr}\MatH\X+\N.  \label{snrVGC}
\end{equation}

In \cite{Ham2009}, for a particular, effectively finite-dimensional vector Gaussian channel, an identity analogous to \eqref{I-MMSE} 
has been derived. There, the probability distribution of the input vector $\X$ depends on $\snr$ such that the mutual 
information occurring in \eqref{I-MMSE} achieves capacity of the vector Gaussian channel; however, the right-hand side (RHS) of that 
identity differs from (half) the MMSE as given in \eqref{I-MMSE}. In \cite{Ham2014}, the same vector Gaussian channel as in \cite{Ham2009} 
arose from a particular continuous-time Gaussian channel through discretization by optimal detection \cite{North} 
of the channel output signals with the use of matched filters, following the approach in \cite{Gallager} for linear 
time-invariant channels; after a certain normalization, the aforementioned RHS has been recognized as (half) the NODE 
(to be defined later) of the channel output signals. In this way, a first instance of the C--NODE relationship has been 
encountered.

The goal of the present paper is to extend this C--NODE relationship 1) to more general vector Gaussian channels, 2) to 
continuous-time Gaussian channels in the form of 
the linear time-varying (LTV) channels considered in \cite{Ham2016}, and 3) to compare the C--NODE relationship with the 
I--MMSE relationship in either case.
\subsubsection*{Notation}
We use natural logarithms and so the unit nat for all information measures. $\mathcal{N}(0,\theta^2)$ is the Gaussian 
distribution with mean 0 and variance $\theta^2$. $\S(\mathbb{R}^2)$ is the Schwartz space of rapidly decreasing 
functions on $\mathbb{R}^2$; $\S_{\ge0}(\mathbb{R}^2)$ is the set of all non-negative real-valued functions in 
$\S(\mathbb{R}^2)$. $x^+$ denotes the positive part of $x\in\mathbb{R}$, $x^+=\max\{0,x\}$. For any two functions 
$A=A(r), B=B(r):[1,\infty)\rightarrow\mathbb{R}$ the notation $A\doteq B$ means $A(r)=B(r)+o(r^2)$ as $r\rightarrow\infty$, 
where $o(\cdot)$ is the standard Landau little-o symbol (cf. \cite{Ham2016}).

\section{Vector Gaussian Channels} \label{Sec_VGC}
\subsection{Detection, Capacity, and Parameter $\snr$} \label{Sec_VGC_A}
Consider the vector Gaussian channel
\begin{equation}
   \Y=\MatH\X+\N,                                                                          \label{VGC}
\end{equation}
where $\MatH$ is a determinstic real $L\times L$ matrix and the noise vector $\N=(N_0,\ldots,N_{L-1})^{\T}$ has 
independent random components $N_k\sim\mathcal{N}(0,\theta^2),\,k=0,\ldots,L-1,$ with the noise variance 
$\theta^2>0$, $\X=(X_0,\ldots,X_{L-1})^{\T}$ is the random input vector, and $\Y=(Y_0,\ldots,Y_{L-1})^{\T}$ the 
corresponding output. If $\x=(x_0,\ldots,x_{L-1})^{\T}$ and $\n=(n_0,\ldots,n_{L-1})^{\T}$ are realizations of the random 
vectors $\X$ and $\N$, resp., then the realization $\y=(y_0,\ldots,y_{L-1})^{\T}$ of $\Y$ is determined by the equation
\begin{equation}
  \y=\MatH\x+\n.                                                                   \label{VGCa}
\end{equation}
$\MatH$ has the singular value decomposition (SVD) $\MatH=\G\DELTA\F^{\T}$ with orthogonal $L\times L$ 
matrices $\F$ and $\G$ and a diagonal matrix $\DELTA=\diag(\sqrt{\lambda_0},\ldots,\sqrt{\lambda_{L-1}})$, where 
$\lambda_0\ge\lambda_1\ge\ldots\ge\lambda_{L-1}\ge0$ are the eigenvalues of $\MatH^{\T}\MatH$ 
(counting multiplicity). Occasionally, we shall use the invertible matrix $\DELTA_\epsilon$ obtained from $\DELTA$ by 
replacing zeros on the diagonal with some $\epsilon>0$. Writing 
$\F=(\f_0\ldots\f_{L-1})$, $\G=(\g_0\ldots\g_{L-1})$ with the column vectors $\f_k,\,\g_k$, it then holds for every 
(column) vector $\x\in\R^L$ that
\[
  \MatH\x=\sum_{k=0}^{L-1} \sqrt{\lambda_k}a_k\,\g_k,
\]
where $a_k=\langle\x,\f_k\rangle\triangleq\x^{\T}\f_k$. Since only the coefficients $a_k$ carry information, 
the linear combination $\x=\sum_{k=0}^{L-1}a_k\f_k$ would be a suitable channel input vector. 
At the receiver, the perturbed vector $\v=\MatH\x$, $\y=\v+\n,$ is passed through a bank of matched filters 
$\langle\,\cdot\,,\g_k\rangle,\,k=0,\ldots,L-1$. The matched filter output signals are $\langle\y,\g_k\rangle=b_k+e_k$, 
where $b_k=\langle \v,\g_k\rangle=\sqrt{\lambda_k}a_k$, and the detection errors 
$e_k=\langle\n,\g_k\rangle=g_{k0}n_0+\ldots+g_{k,L-1}n_{L-1}$ are realizations of independent 
identically distributed Gaussian random variables $E_k\sim\mathcal{N}(0,\theta^2)$. From the detected values 
$\hat{b}_k=b_k+e_k$ we get the estimates $\hat{a}_k=\hat{b}_k/\sqrt{\lambda_k}=a_k+z_k$ of the coefficients $a_k$ for 
the input vector $\x$, where $z_k$ are realizations of independent Gaussian random variables 
$Z_k\sim\mathcal{N}(0,\theta^2/\lambda_k)$ (put $\theta^2/0=\infty$). Thus, we are led to the new vector Gaussian channel
\begin{equation}
  \Y=\X+\Z,  \label{VGC_equiv}
\end{equation}
where the random components $Z_k$ of the noise vector $\Z=(Z_0,\ldots,Z_{L-1})^{\T}$ are distributed as described. 
The vector Gaussian channels \eqref{VGC} and \eqref{VGC_equiv} are equivalent in the sense that for any average input energy $S$ their 
capacity $C(S)$ is the same. Indeed, since mutual information is invariant with respect to invertible linear 
transformations, we have
\begin{align}
  I(\X;\X+\Z)&=\lim_{\epsilon\rightarrow 0}I(\X;\X+\DELTA_\epsilon^{-1}\N) \nonumber\\
             &=\lim_{\epsilon\rightarrow 0}I(\X;\DELTA_\epsilon\X+\N) \nonumber\\
	     &=I(\X;\DELTA\X+\N) \label{EqMutI}\\
	     &= I(\tilde{\X};\MatH\tilde{\X}+\tilde{\N}),\nonumber
\end{align}
where $\tilde{\X}=\F\X$ is an arbitrary vector with the property that 
$\|\tilde{\X}\|\triangleq\langle\tilde{\X},\tilde{\X}\rangle^{1/2}=\|\X\|$, and $\tilde{\N}=\G\N$ has independent 
components $\sim\mathcal{N}(0,\theta^2)$ (as $\N$); consequently, 
\[
  C(S)=\max_{\E\|\X\|^2\le S}I(\X;\X+\Z)=\max_{\E\|\tilde{\X}\|^2\le S}I(\tilde{\X};\MatH\tilde{\X}+\tilde{\N}).
\]

The capacity of the vector Gaussian channel \eqref{VGC_equiv} is computed by waterfilling on the noise variances 
\cite[Th.~7.5.1]{Gallager}. Let $\nu_k^2=\theta^2/\lambda_k,\,k=0,1,\ldots,L-1,$ be the noise variance in the 
$(k+1)$st subchannel of the channel \eqref{VGC_equiv}. Precluding the trivial case $S=0$, the ``water level" $\sigma^2$ 
is then uniquely determined by the condition
\begin{equation}
  S = \sum_{k=0}^{K-1} (\sigma^2-\nu_k^2)=\sum_{k=0}^{L-1} (\sigma^2-\nu_k^2)^+,\label{def_sigma}
\end{equation}
where $K=\max\{k\in\mathbb{N};\,\nu_{k-1}^2<\sigma^2,\,k\le L\}$ is the number of active subchannels. The capacity 
$C(S)$ is achieved, if the input vector $\X=(X_0,\ldots,X_{L-1})^{\T}$ has independent components 
$X_k\sim \mathcal{N}(0,\sigma^2-\nu_k^2)$ for $k=0,\ldots,K-1$ and $X_k=0$ else; then
\[
  C(S)=\sum_{k=0}^{K-1}\frac{1}{2}\ln\left(1+\frac{\sigma^2-\nu_k^2}{\nu_k^2}\right)
      =\frac{1}{2}\sum_{k=0}^{K-1}\ln(\snr\,\lambda_k),
\]
where $\snr\triangleq\sigma^2/\theta^2$ is the signal-to-noise ratio. Since always $\snr\,\lambda_0=\sigma^2/\nu_0^2\ge1,$ 
$\lambda_0^{-1}$ is the smallest feasible $\snr$ (assumed when $S=0$).
\begin{remark} Since, in the case of $\lambda_0=1$, only the portion $\sigma^2-\theta^2$ contributes to the signal, 
$\sigma^2/\theta^2$ is, then, rather a \textit{signal plus noise}-to-noise ratio; we stick to the notation ``snr" to 
conform with \cite{GSV}.
\end{remark}

\subsection{C--NODE Relationship in Vector Gaussian Channels}
Because of Eq.~\eqref{EqMutI}, for the channel \eqref{VGC_equiv} it holds that
\[
  I(\X;\Y) = I\left(\X';\frac{\sigma}{\theta}\DELTA\X'+\N'\right),
\]
where $\X'=\sigma^{-1}\X$, and $\N'=\theta^{-1}\N$ has independent standard Gaussian components, $\N$ being the noise 
vector in \eqref{VGC}. Capacity is achieved, if $\X'=(X_0',\ldots,X_{L-1}')^{\T}$ has independent components 
$X_k'\sim \mathcal{N}(0,1-\snr^{-1}\lambda_k^{-1})$ for $k=0,\ldots,K-1$ and $X_k'=0$ else. Putting $\X''=\F\X'$, 
we obtain
\begin{equation}
  I(\X';\sqrt{\snr}\DELTA\X'+\N') = I(\X'';\sqrt{\snr}\,\MatH\X''+\N''),  \label{I-snr}
\end{equation}
where $\N''=\G\N'$ has independent standard Gaussian components, independent of $\X''$. Capacity is achieved, if $\X''=\F\X'$ 
where $\X'$ is distributed as above. Since the RHS of Eq.~\eqref{I-snr} then only depends on $\snr$, 
we may write (with slight abuse of notation)
\begin{align}
  C(\snr) &= I(\X'';\sqrt{\snr}\,\MatH\X''+\N'') \label{C1-VGC}\\
          &= \frac{1}{2}\sum_{k=0}^{K-1}\ln(\snr\,\lambda_k). \label{C2-VGC}
\end{align}
The RHS of Eq.~\eqref{C1-VGC} is reminescent of the mutual information occurring in the I--MMSE relationship 
\eqref{I-MMSE}. It is therefore tempting to take the derivative of the RHS of Eq.~\eqref{C2-VGC} with respect to $\snr$. 
Before doing so, observe that $K$ depends on $\snr$ since $K=K(\snr)=\max\{k\in\mathbb{N};\,\lambda_{k-1}>\snr^{-1},k\le L\}$ 
(put $\max\emptyset=0$); on the other hand, $K(\snr)$ is piecewise constant. Excluding those $\snr$'s where $K(\snr)$ makes a 
jump, we thus obtain
\[
  \frac{\d}{\d\,\snr}C(\snr)=\frac{1}{2}\frac{K}{\snr}=\frac{1}{2}\frac{K\theta^2}{\sigma^2}=\frac{1}{2}K(\theta/\sigma)^2.
\]

Due to the application of matched filters for detection, \emph{optimal} detection has been performed \cite{North}. Therefore, 
$K\theta^2$ is the (total) optimal detection error; division by $\sigma^2$ may be regarded as a normalization. Likewise, 
after normalization $\sigma^2\rightarrow 1,\,\theta^2\rightarrow(\theta/\sigma)^2$ (retaining the $\snr$), 
$K(\theta/\sigma)^2$ would be the (total) optimal detection error. Anyway, the following definition appears appropriate:
\begin{definition} \label{Def_NODE} For any feasible $\snr$, the NODE in the vector Gaussian channel \eqref{VGC} is given 
by the function
\[
  \node(\snr)=\frac{K(\snr)}{\snr}=K(\snr)\,\theta'^2,
\]
where $\theta'^2\triangleq(\theta/\sigma)^2$ is called the \textit{primitive} NODE. If $\snr$ is infeasible, we put 
formally $\node(\snr)=0$.
\end{definition}

Note that for any feasible $\snr$, $\snr^{-1}$ is identical to the primitive NODE $\theta'^2$; so, no further normalization is needed 
when working with $\snr^{-1}$ (or $\snr$) in the feasible case.
\begin{theorem} For all $\snr>\lambda_0^{-1}$ but for at most $L-1$ exceptions, the capacity $C(\snr)$ of the vector Gaussian channel \eqref{VGC} 
is differentiable and satisfies
\begin{equation}
  \frac{\d}{\d\,\snr}C(\snr)=\frac{1}{2}\node(\snr).  \label{C-NODE}
\end{equation}
\end{theorem}
\begin{IEEEproof} For growing $\snr$, differentiability breaks down when a new subchannel is added. This occurs as soon as 
$\lambda_K$ ($K$ being the actual number of subchannels) exceeds $\snr^{-1}$, which happens at most $L-1$ times. The rest 
of the theorem has already been proved.
\end{IEEEproof}

We observe a striking similarity between the I--MMSE relationship \eqref{I-MMSE} and the C--NODE relationship \eqref{C-NODE}. 
Note that the part of the estimation error in \eqref{I-MMSE} is taken by a \emph{detection} error in \eqref{C-NODE}.

\subsection{Comparison of the NODE With the MMSE in Vector Gaussian Channels} \label{NODEvsMMSE_1}
To understand the difference between Eq.~\eqref{I-MMSE} and Eq.~\eqref{C-NODE} in more detail, we calculate the 
MMSE.\footnote{For simplicity, we assume that the diagonal matrix $\DELTA$ in the SVD of $\MatH$ is invertible and 
that the number $K$ of active subchannels is equal to $L$ (both assumptions can be removed).} Following \cite{GSV}, given
\begin{equation} \label{MIMO_GSV}
  \Y=\sqrt{\snr}\MatH\X''+\N'',
\end{equation}
the MMSE in estimating $\MatH\X''$ is
\begin{align*}
  \mmse(\snr)&=\E\left\|\MatH\X''-\MatH\widehat{\X''}\right\|^2\\
             &=\tr[\MatH(\SIGMA''^{-1}+\snr\,\MatH^{\T}\MatH)^{-1}\MatH^{\T}],
\end{align*}
where $\widehat{\X''}$ is the minimum mean-square estimate of $\X''$, and $\SIGMA''^{-1}$ is the inverse of the covariance 
matrix $\SIGMA''$ of $\X''$ which is given by
\[
  \SIGMA''=\E[\X''\X''^{\T}]=\E[\F\X'\X'^{\T}\F^{-1}]=\F\SIGMA'\F^{-1},
\]
where $\SIGMA'=\E[\X'\X'^{\T}]$ is the covariance matrix of $\X'$. If $\X'$ has independent Gaussian components 
$X_k'\sim\mathcal{N}(0,1-\snr^{-1}\lambda_k^{-1}),\,k=0,\ldots,K-1,$ and $K=L$ (as assumed), then 
$\SIGMA'=\diag(1-\snr^{-1}\lambda_k^{-1},\,k=0,\ldots,K-1)$. We now compute
\begin{align*}
  \mmse(\snr)=&\;\tr[\G\DELTA\F^{-1}(\F\SIGMA'^{-1}\F^{-1}\!+\snr\,\F\DELTA^2\F^{-1})^{-1}\\
              &\;\cdot\,\F\DELTA\G^{-1}]\\
	     =&\;\tr[\G\DELTA(\SIGMA'^{-1}+\snr\,\DELTA^2)^{-1}\DELTA\G^{-1}]\\
	     =&\;\tr\{[(\DELTA^2\SIGMA')^{-1}+\snr\,I_K]^{-1}\},
\end{align*}
where $I_K$ is the $K\times K$ identity matrix. So, we obtain
\begin{align}
  \mmse(\snr) 
        &=\sum_{k=0}^{K-1}\snr^{-1}(1-\snr^{-1}\lambda_k^{-1})  \label{VGC-MMSE}\\
	&=\frac{K}{\snr}-\snr^{-1}\sum_{k=0}^{K-1}\snr^{-1}\lambda_k^{-1}, \nonumber
\end{align}
or, equivalently,
\begin{align}
  \node(\snr)&=\mmse(\snr)+\left\{\sum_{k=0}^{K-1}\snr^{-1}\lambda_k^{-1}\right\}\snr^{-1} \label{NODE-MMSE_1}\\
             &>\mmse(\snr)  \label{NODE-MMSE_2}.  
\end{align}
\begin{remark} Similarly to \cite[Sec.\ II-D.2]{GSV}, it can be shown that the expression in curly brackets $\{\ldots\}$ 
in Eq.~\eqref{NODE-MMSE_1} is the trace of a Fisher information matrix.
\end{remark}

Now, the strict inequality \eqref{NODE-MMSE_2} prompts the following observation: The increase of capacity with growing $\snr$ 
as given by Eq.~\eqref{C-NODE} is always larger than anticipated by the I--MMSE relationship \eqref{I-MMSE}. The resolution 
of this seeming contradiction is the implicit assumption in \cite[Th.~2]{GSV} that the probability distribution of the 
channel input vector $\X$ does not depend on $\snr$. Refer to \cite{Bustin} concerning possible extensions 
of the I--MMSE relationship to the $\snr$-dependent case.

\section{Continuous-Time Gaussian Channels}
\subsection{Channel Model and Discretization}
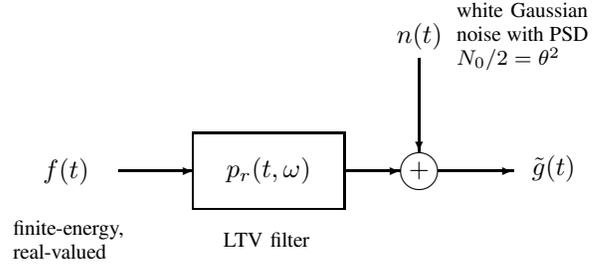
\begin{figure}
\setlength{\unitlength}{1cm}
\begin{picture}(8.89,4)
  \put(3,1){\framebox(2,1){$p_r(t,\omega)$}}
  \put(2,1.5){\vector(1,0){1}}
  \put(1.0,1.4){$f(t)$}
  \put(6.25,1.5){\vector(1,0){1.0}}
  \put(5.875,1.375){\makebox(0.25,0.25){$+$}}
  \put(5,1.5){\vector(1,0){0.75}}
  \put(7.5,1.45){$\tilde{g}(t)$}
  \put(6,3){\vector(0,-1){1.25}}
  \put(5.7,3.2){$n(t)$}
  \put(6,1.5){\circle{0.5}}
  \put(6.5,3.2){\parbox{1.75cm}{\footnotesize white Gaussian noise with PSD $N_0/2=\theta^2$}}
  \put(0.6,0.5){\parbox{1.75cm}{\footnotesize finite-energy, real-valued}}
  \put(3.4,0.5){\parbox{2cm}{\footnotesize LTV filter}}
\end{picture}
\caption{Model of the continuous-time Gaussian channel.} \label{Fig_1}
\end{figure}

Consider as in \cite{Ham2016} for any spreading factor $r\ge 1$ held constant the LTV channel
\begin{equation}
   \tilde{g}(t)=(\P _rf)(t)+n(t),\,-\infty<t<\infty,  \label{LTV_Ch}
\end{equation}
where $\P _r:L^2(\mathbb{R})\rightarrow L^2(\mathbb{R})$ is the LTV filter (operator) with the spread Weyl symbol 
$p_r(t,\omega)\triangleq p(t/r,\omega/r),\,p\in\S(\mathbb{R}^2)$;  the kernel $h(t,t')$ 
of operator $\P=\P_1$ is assumed to be real-valued.  The real-valued filter input signals $f(t)$ are of finite energy 
and the noise signals  $n(t)$ at the filter output are realizations of white Gaussian noise with two-sided 
power spectral density (PSD) $N_0/2=\theta^2>0$. This channel is depicted in Fig.~\ref{Fig_1}. As in 
\cite{Ham2016}, it may be assumed that the operator $\P_r\!\!^*\P_r$ has infinitely many eigenvalues 
$\lambda_0^{(r)}\ge\lambda_1^{(r)}\ge\ldots\ge 0$ (counting multiplicity) and that $\lambda_k^{(r)}\rightarrow 0$ 
as $k\rightarrow\infty$. As shown in \cite[Sec.~III]{Ham2016}, optimal detection by means of matched filters then 
leads to the infinite-dimensional vector Gaussian channel
\begin{equation}
  Y_k=X_k+Z_k,\,Z_k\sim\mathcal{N}(0,\theta^2/\lambda_k^{(r)}),\,k=0,1,\ldots,   \label{inftyVGC}
\end{equation}
where the noise $Z_k$ is independent from subchannel to subchannel.

\subsection{C--NODE Relationship in Continuous-Time Gaussian Channels} 
From the waterfilling theorem \cite[Th.~2]{Ham2016} we know that under a quadratic growth condition imposed on 
the average input energy $S=S(r)$, the capacity of the LTV channel \eqref{LTV_Ch} is given with the use of the 
``cup" function
\[
  N_r(t,\omega)=\frac{\theta^2}{2\pi}|p_r(t,\omega)|^{-2}
\]
by the equation
\begin{equation}
   C\doteq \frac{1}{2\pi}\iint_{\mathbb{R}^2}\frac{1}{2}
       \ln\left(1+\frac{(\nu-N_r(t,\omega))^+}{N_r(t,\omega)}\right)\,\d t\,\d\omega,   \label{C}
\end{equation}
where the ``water level" $\nu$ is chosen so that
\begin{equation}
   S\doteq\iint_{\mathbb{R}^2}(\nu-N_r(t,\omega))^+\,\d t\,\d\omega. \label{S}
\end{equation}

Eq.~\eqref{S} has been derived in \cite{Ham2016} from the original waterfilling condition
\[
  S(r)=\sum_{k=0}^{K-1} (\sigma^2-\nu_k^2(r))=\sum_{k=0}^\infty (\sigma^2-\nu_k^2(r))^+,
\]
where $\nu_k^2(r)=\theta^2/\lambda_k^{(r)},k=0,1,\ldots,$ are the noise variances and $\sigma^2=2\pi\nu$. In the 
present context, $\sigma^2=\snr\,\theta^2$ so that $\sigma^2$ does not depend on $r$ (and the quadratic growth 
condition imposed on $S$ is automatically fulfilled); the number $K$ of active subchannels depends on $r$ and $\snr$ 
since
\[
  K=K(r,\snr)=\max\{k\in\mathbb{N};\lambda_{k-1}^{(r)}>\snr^{-1}\},
\]
again putting $\max\emptyset=0$.
\begin{theorem} \label{Th1} For any fixed $\snr>0$ it holds that
\begin{equation}
  K(r,\snr)\doteq\check{K}(r,\snr)\triangleq\frac{1}{2\pi}\,
                           \iint\limits_{|p_r(t,\omega)|^2\ge\snr^{-1}}1\,\d t\,\d\omega.\label{K}
\end{equation}
\end{theorem}
\begin{IEEEproof}
With the use of the (modified) Heaviside function
\[
  H(x)=\left\{\!\!\begin{array}{ll} 0, & \mbox{if }x\le 0,\\
                                    1, & \mbox{if }x>0
                  \end{array}\right.
\]
we can write
\[
  K(r,\snr)=\sum_{k=0}^\infty H\left(1-\frac{\nu_k^2(r)}{\sigma^2}\right).
\]
For $\delta\in(0,1)$, replace $H(x)$ with the continuous function
\[
  H_\delta(x)=\left\{\!\!\begin{array}{ll} 0, & \mbox{if }x\le 0,\\
                                \delta^{-1}x, & \mbox{if }0<x<\delta,\\
				1, & \mbox{if }x\ge\delta
                  \end{array}\right.
\]
and define $H_{-\delta}(x)=H_\delta(x+\delta),\,x\in\mathbb{R}$. Putting
\[
  K_{\pm\delta}(r,\snr)=\sum_{k=0}^\infty H_{\pm\delta}\left(1-\frac{\nu_k^2(r)}{\sigma^2}\right),
\]
we then obtain
\begin{equation}
  K_\delta(r,\snr)\le K(r,\snr)\le K_{-\delta}(r,\snr).  \label{Eq_Kpmdelta}
\end{equation}
Since
\begin{align*}
  K_\delta(r,\snr)&=\sum_{k=0}^\infty H_\delta\left(1-\frac{1}{\snr\,\lambda_k^{(r)}}\right)\\
                  &=\sum_{k=0}^\infty a(r)g(b(r)\lambda_k^{(r)}),
\end{align*}
where $a(r)=1,\,b(r)=\snr$ and
\[
  g(x)=\left\{\!\!\begin{array}{ll} H_\delta\left(1-\frac{1}{x}\right), & \mbox{if }x> 0,\\
                                    0, & \mbox{if }x=0,
                  \end{array}\right.
\]
the Szeg\H{o} theorem \cite[Th. 1]{Ham2016} applies and yields
\begin{align*}
  K_\delta(r,\snr)&\doteq \frac{1}{2\pi}\iint H_\delta\left(1-\frac{1}{\snr\,|p_r(t,\omega)|^2}
                                                                           \right)\,\d t\,\d\omega \\
	&=\check{K}(r,\snr)-I_\delta(r,\snr),
\end{align*}
that is,
\[
  K_\delta(r,\snr)/r^2=(\check{K}(r,\snr)-I_\delta(r,\snr))/r^2+\epsilon_1,
\]
where $\epsilon_1\rightarrow 0$ as $r\rightarrow\infty$. For $I_\delta(r,\snr)$ it is readily seen that 
\[
  0\le I_\delta(r,\snr)\le \frac{r^2}{2\pi}\,\iint\limits_{\frac{1}{\snr}<|p(t,\omega)|^2
                                       <\frac{1}{(1-\delta)\snr}} 1\,\,\d t\,\d\omega = \epsilon_2r^2,
\]
where $\epsilon_2\rightarrow 0$ as $\delta\rightarrow 0$. Therefore, 
$K_\delta(r,\snr)/r^2=\check{K}(r,\snr)/r^2+\epsilon$, where $\epsilon\rightarrow 0$ if $\delta$ becomes 
arbitrarily small and, then, $r\rightarrow\infty$; a similar result is obtained for 
$K_{-\delta}(r,\snr)$. In combination with Ineq.~\eqref{Eq_Kpmdelta}, this proves the theorem.
\end{IEEEproof}

The RHS of Eq.~\eqref{C} reduces to the double integral
\begin{equation}
  \check{C}(r,\snr)\triangleq\frac{1}{4\pi}\iint\limits_{|p_r(t,\omega)|^2\ge\snr^{-1}}
                                \ln(\snr\,|p_r(t,\omega)|^2)\,\d t\,\d\omega.  \label{Cbreve}
\end{equation}
We say that a function $u\in\S_{\ge0}(\mathbb{R}^2)$ is \emph{non-flat}, if for every constant $c>0$ the Lebesgue measure 
(or area) of the level curve $\{(x,y);u(x,y)=c\}$ is zero (no assumption is made about the area 
of the set $\{(x,y);u(x,y)=0\}$).
\begin{lemma} \label{Lm1}
Let $u\in\S_{\ge0}(\mathbb{R}^2)$ be non-flat. Define for all $s>0$ the function
\[
  J(s)=\iint\limits_{u(x,y)\ge s^{-1}}\ln(s u(x,y))\,\d x\,\d y.
\]
Then for all $s>0$ it holds that
\begin{equation}
  J'(s)=\frac{1}{s}\iint\limits_{u(x,y)\ge s^{-1}}1\,\d x\,\d y.  \label{Jprimes}
\end{equation}
\end{lemma}
\begin{IEEEproof}
We consider only the region $\Omega_s$ (see Fig.~\ref{Fig_2}) in the first quadrant enclosed 
by the coordinate axes and the boundary line $B_s=\{(x,y);\,u(x,y)=s^{-1},\,0\le x\le x_0,\,y\ge0\}$. Additionally, we 
assume that $B_s$ has the representation $y=y(s,x),\,0\le x\le x_0$.
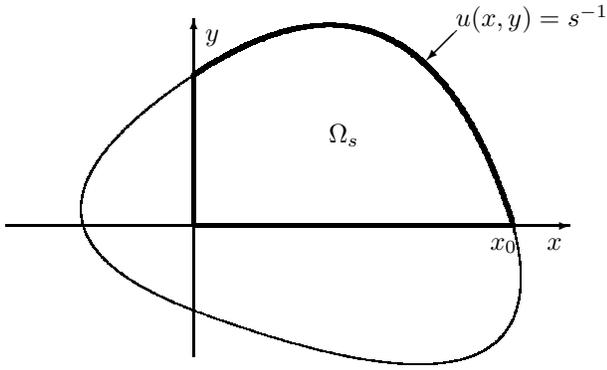
\begin{figure}
\setlength{\unitlength}{1cm}
\begin{picture}(8.89,5)
  \put(4.5,2.7){\makebox(1,1){$\Omega_s$}}
  \put(7.55,1.5){\makebox(0.5,0.5){$x$}}
  \put(3,4.25){\makebox(0.5,0.5){$y$}}
  \put(0.5,2){\vector(1,0){7.5}}
  \put(3,0.25){\vector(0,1){4.5}}
  \put(6.5,4.6){\vector(-1,-1){0.4}}
  \put(6.5,4.25){\makebox(2,1){$u(x,y)=s^{-1}$}}
  \put(6.87,1.48){\makebox(0.5,0.5){$x_0$}}
  \put(7.25,2){\line(0,-1){0.1}}
  \qbezier(3,4)(0,2)(3,0.87)
  \qbezier(3,0.87)(8,-1)(7.25,2)  
  \linethickness{1.5pt}
  \put(3,2){\line(1,0){4.25}}
  \put(3,2){\line(0,1){2}}
  \qbezier(3,4)(6,6)(7.25,2)
\end{picture}
\caption{Sketch for the proof of Lemma~\ref{Lm1}. The region within the closed line is the set of all points 
$(x,y)\in\mathbb{R}^2$ with the property that $u(x,y)\ge s^{-1}$.} \label{Fig_2}
\end{figure}
For the integral
\begin{align*}
  J_1(s)&=\iint_{\Omega_s}\ln(s u(x,y))\,\d x\,\d y\\
        &=\int_0^{x_0}\int_0^{y(s,x)}\ln(s u(x,y))\,\d y\,\d x
\end{align*}
we get by differentiation that
\begin{align*}
  J_1'(s)=&\int_0^{x_0}\int_0^{y(s,x)}\partial\ln(s u(x,y))/\partial s\,\d y\,\d x\\
         &+\int_0^{x_0}\frac{\partial y}{\partial s}(s,x) \ln(su(x,y(s,x)))\,\d x\\
	 =&\frac{1}{s}\iint_{\Omega_s}1\,\d x\,\d y+0.
\end{align*}
A generalization to the other quadrants and other boundary geometries should now be obvious. Addition of the separate 
double integrals yields Eq.~\eqref{Jprimes} and completes the proof.
\end{IEEEproof}

For ease of presentation, we assume from now on that the squared absolute value of the Weyl symbol 
$p$ of $\P$ is non-flat.
\begin{theorem}For any fixed $\snr>0$, the approximate capacity \eqref{Cbreve} satisfies 
\begin{equation}
  \frac{\d}{\d\,\snr}\check{C}(r,\snr)\doteq \frac{1}{2} \node(r,\snr), \label{Diff_Csnr}
\end{equation}
where the NODE is given by
\begin{equation}
  \node(r,\snr)\doteq \frac{\snr^{-1}}{2\pi}\,\iint
                     \limits_{|p_r(t,\omega)|^2\ge\snr^{-1}}1\,\d t\,\d\omega.  \label{Eq_intnode}
\end{equation}
\end{theorem}
\begin{IEEEproof}
By means of Lemma~\ref{Lm1} and Theorem~\ref{Th1} we get in combination with Definition~\ref{Def_NODE} [generalized 
to the $r$-dependent, effectively finite-dimensional vector Gaussian channel \eqref{inftyVGC}] that
\begin{align*}
  \frac{\d}{\d\,\snr}\check{C}(r,\snr)&=\frac{1}{2}\cdot\frac{1}{\snr}\check{K}(r,\snr)\\
                          &\doteq\frac{1}{2}\cdot\frac{1}{\snr}\cdot K(r,\snr)\\
			  &=\frac{1}{2}\node(r,\snr),
\end{align*}
which proves both Eq.~\eqref{Diff_Csnr} and Eq.~\eqref{Eq_intnode}.
\end{IEEEproof}

\subsection{Comparison of the NODE With the MMSE in Continuous-Time Gaussian Channels}
If $r\ge 1$ and $\snr\ge 0$ are held constant, then $K=K(r,\snr)$ is finite so that Eq.~\eqref{VGC-MMSE} 
for the MMSE carries over to the infinite-dimensional setting \eqref{inftyVGC} without changes and yields
\begin{align*}
  \mmse(r,\snr)&=\sum_{k=0}^{K-1}\frac{1}{\snr}\left(1-\frac{1}{\snr\,\lambda_k^{(r)}}\right) \nonumber\\
         &=\sum_{k=0}^\infty\frac{1}{\snr}\left(1-\frac{1}{\snr\,\lambda_k^{(r)}}\right)^+.
\end{align*}
 Recalling that $p\in\S(\mathbb{R}^2)$, the Szeg\H{o} theorem \cite[Th. 1]{Ham2016} may be applied to 
 the last expression; so we continue
\begin{align}
  \lefteqn{\mmse(r,\snr)} \nonumber\\
     &\doteq\frac{1}{2\pi}\iint\frac{1}{\snr}\left(1-\frac{1}{\snr\,|p_r(t,\omega)|^2}
                                                                 \right)^+\,\d t\,\d\omega \label{Eq_intmmse}\\
     &=\frac{\snr^{-1}}{2\pi}\iint\limits_{|p_r(t,\omega)|^2\ge\snr^{-1}}
          \!\!\!\!\left(1-\frac{1}{\snr\,|p_r(t,\omega)|^2}\right)\,\d t\,\d\omega \nonumber\\
     &<\frac{\snr^{-1}}{2\pi}\iint\limits_{|p_r(t,\omega)|^2\ge\snr^{-1}}1\,\d t\,\d\omega \nonumber\\
     &\doteq\node(r,\snr). \nonumber      
\end{align}
In order to get rid of the error term $o(r^2)$ involved in the dotted equations \eqref{Eq_intnode} and 
\eqref{Eq_intmmse}, we average with respect to $r^2$ and obtain
\begin{align*}
  \overline{\node}(\snr)&\triangleq \lim_{r\rightarrow\infty}\frac{\node(r,\snr)}{r^2}
       =\frac{\snr^{-1}}{2\pi}\iint\limits_{|p(t,\omega)|^2\ge\frac{1}{\snr}}1\,\d t\,\d\omega,\\
  \overline{\mmse}(\snr)&\triangleq \lim_{r\rightarrow\infty}\frac{\mmse(r,\snr)}{r^2}\\
       &=\frac{\snr^{-1}}{2\pi}\iint\limits_{|p(t,\omega)|^2\ge\frac{1}{\snr}}
                     \!\!\!\!\left(1-\frac{1}{\snr\,|p(t,\omega)|^2}\right)\,\d t\,\d\omega.
\end{align*}
Thus, for all $\snr>M^{-1},\,M=\max_{t,\omega}|p(t,\omega)|^2,$ it holds the strict inequality
\[
  \overline{\node}(\snr)>\overline{\mmse}(\snr),
\]
which is similar to Ineq.~\eqref{NODE-MMSE_2} for $\node(\snr)$ and $\mmse(\snr)$ in finite-dimensional vector 
Gaussian channels. In the case of $0\le\snr<M^{-1}$ it holds, of course, that
\[
  \overline{\node}(\snr)=0=\overline{\mmse}(\snr).
\]

\begin{example} \label{Ex_LTV} Consider the operator $\P_r:L^2(\mathbb{R})\rightarrow L^2(\mathbb{R})$, $r\ge1$,  
with the bivariate Gaussian function
\[
  p_r(t,\omega)=\mathrm{e}^{-\frac{1}{2r^2}(\gamma^{-2}t^2+\gamma^2\omega^2)},
\]
$\gamma>0$ fixed, as the (spread) Weyl symbol. Here we have $M=\max_{t,\omega}|p_1(t,\omega)|^2=1$. Computation of the RHS 
of Eq.~\eqref{Eq_intnode} yields for any $\snr\ge M^{-1}=1$ held constant the equation
\[
  \node(r,\snr)\doteq\frac{r^2}{2}\frac{\ln\snr}{\snr}.
\]
In virtue of the C--NODE relationship~\eqref{Diff_Csnr} we obtain from the foregoing NODE by integration the 
capacity
\[
  C(r,\snr)\doteq\frac{r^2}{8}(\ln\snr)^2,
\]
which indeed coincides with the capacity directly obtained from Eq.~\eqref{C} (expressed as a function of $r$ and 
$\snr$).

Further, computation of the RHS of Eq.~\eqref{Eq_intmmse} gives
\[
  \mmse(r,\snr)\doteq\frac{r^2}{2}\left\{\frac{\ln\snr}{\snr}-\frac{1}{\snr}\left(1-\frac{1}{\snr}\right)\right\}.
\]
Averaging with respect to $r^2$ as $r\rightarrow\infty$ finally yields
\begin{gather*}
   \overline{\node}(\snr)=\frac{1}{2}\frac{\ln\snr}{\snr},\\
   \overline{\mmse}(\snr)=\overline{\node}(\snr)-\frac{1}{2\,\snr}\left(1-\frac{1}{\snr}\right).
\end{gather*}
\begin{figure}
\centering
\includegraphics[width=3.5in]{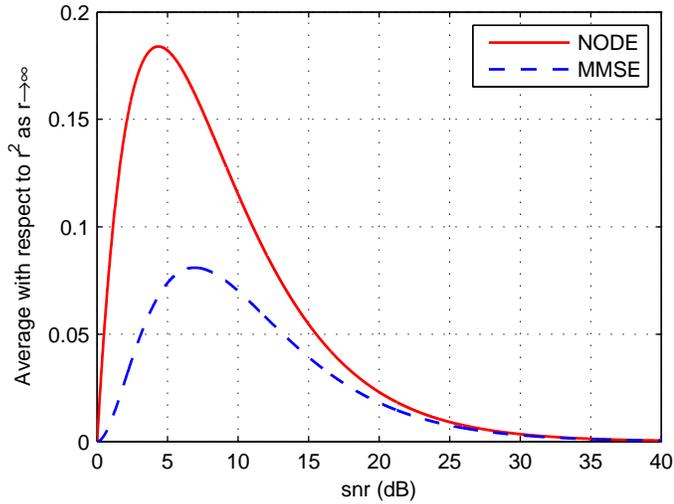}
\caption{Comparison of the NODE with the MMSE in the continuous-time Gaussian channel of Example~\ref{Ex_LTV}.}
\label{Figure_3}
\end{figure}

In Fig.~\ref{Figure_3}, $\overline{\node}(\snr)$ and $\overline{\mmse}(\snr)$ are plotted against $10\log_{10}\snr$ 
for $\snr\ge1$. Observe the difference in size.
\end{example}



\begin{thebibliography}{10}
\bibitem{GSV}
D.~Guo, S.~Shamai~(Shitz), and S.~Verd\'{u}, ``Mutual information and minimum mean-square error in Gaussian channels," \emph{IEEE 
Trans. Inf. Theory}, vol.~51, pp.~1261--1282, 2005.
\bibitem{Ham2009}
E.~Hammerich, ``On the heat channel and its capacity," in \emph{Proc.\ IEEE Int.\ 
Symp.\ Inf.\ Theory}, Seoul, South Korea, 2009, pp.~1809--1813.
\bibitem{Ham2014}
E.~Hammerich, ``On the capacity of the heat channel, waterfilling in the time-frequency plane, 
and a C-NODE relationship," 2014 [Online]. Available: arXiv:1101.0287
\bibitem{North}
D.~O.~North, \emph{Analysis of the factors which determine signal/noise discrimination in pulsed-carrier 
systems}. Rept. PTR-6C, RCA Labs., Princeton, NJ, USA, 1943. 
\bibitem{Gallager}
R.~G.~Gallager, \emph{Information Theory and Reliable Communication}. 
New York, NY:\ Wiley, 1968.
\bibitem{Ham2016}
E.~Hammerich, ``Waterfilling theorems for linear time-varying channels and related nonstationary 
sources," \emph{IEEE Trans. Inf. Theory}, vol.~62, pp.~6904--6916, 2016.
\bibitem{Bustin}
R.~Bustin, M.~Payaro, D.~P.~Palomar, and S.~Shamai~(Shitz), ``On MMSE crossing properties and implications in parallel vector 
Gaussian channels," \emph{IEEE Trans. Inf. Theory}, vol.\ 59, pp.~818--844, 2013.
\bibitem{Ham2017}
E.~Hammerich, ``Capacity and normalized optimal detection error in Gaussian channels," 2017 [Online]. Available: arXiv:1701.05523
\end{thebibliography}
\end{document}